# Beam test, simulation, and performance evaluation of PbF$_2$ and PWO-UF crystals with SiPM readout for a semi-homogeneous calorimeter prototype with longitudinal segmentation

C. Cantone [1], S. Carsi [3,4], S. Ceravolo [1], E. Di Meco [1,2], E. Diociaiuti [1],
I. Frank [6,12], S. Kholodenko [5], S. Martellotti [1], M. Mirra [7], P. Monti-Guarnieri [3,4],
M. Moulson [1], D. Paesani [1,2,*], M. Prest [3,4], M. Romagnoni [8], I. Sarra [1],
F. Sgarbossa [9,10], M. Soldani [10,11], E. Vallazza [4]

[1] *INFN Laboratori Nazionali di Frascati, 00044 Frascati RM, Italy*
[2] *Dipartimento di Fisica, Università degli Studi di Roma Tor Vergata, 00133 Roma, Italy*
[3] *Dipartimento di Scienza e Alta Tecnologia, Università degli Studi dell'Insubria, 22100 Como, Italy*
[4] *INFN Sezione di Milano Bicocca, 20126 Milano, Italy*
[5] *INFN Sezione di Pisa, 56100 Pisa, Italy*
[6] *Faculty of Physics, Ludwig Maximilian University of Munich, 80539 Munich, Germany*
[7] *INFN Sezione di Napoli, 80126 Napoli, Italy*
[8] *INFN Sezione di Ferrara, 44122 Ferrara, Italy*
[9] *Dipartimento di Fisica e Astronomia, Università degli Studi di Padova, 35131 Padova, Italy*
[10] *INFN Laboratori Nazionali di Legnaro, 35020 Legnaro PD, Italy*
[11] *Dipartimento di Fisica e Scienze della Terra, Università degli Studi di Ferrara, 44122 Ferrara, Italy*
[12] *CERN, 1211 Geneva 23, Switzerland*

Correspondence*:
D. Paesani
daniele.paesani@lnf.infn.it

**ABSTRACT**

Crilin (Crystal Calorimeter with Longitudinal Information) is a semi-homogeneous, longitudinally segmented electromagnetic calorimeter based on high-$Z$, ultra-fast crystals with UV-extended SiPM readout. The Crilin design has been proposed as a candidate solution for both a future Muon Collider barrel ECAL and for the Small Angle Calorimeter of the HIKE experiment. As a part of the Crilin development program, we have carried out beam tests of small ($10 \times 10 \times 40$ mm$^3$) lead fluoride (PbF$_2$) and ultra-fast lead tungstate (PbWO$_4$, PWO-UF) crystals with 120 GeV electrons at the CERN SPS to study the light yield, timing response, and systematics of light collection with a proposed readout scheme. For a single crystal of PbF$_2$, corresponding to a single Crilin cell, a time resolution of better than 25 ps is obtained for >3 GeV of deposited energy. For a single cell of PWO-UF, a time resolution of better than 45 ps is obtained for the same range of deposited energy. This timing performance fully satisfies the design requirements for the Muon Collider and HIKE experiments. Further optimizations of the readout scheme and crystal surface preparation are expected to bring further improvements.

# 1 INTRODUCTION

Calorimetry for future experiments will require novel solutions to meet the challenges posed by the next generation of high-energy physics experiments carried out at higher and higher intensities. An innovative approach for facing these challenges is represented by the Crystal Calorimeter with Longitudinal Information (Crilin) concept. Crilin is a semi-homogeneous electromagnetic calorimeter with longitudinal segmentation, composed by stackable and interchangeable modules housing high-granularity crystal matrices readout by UV-extended, surface-mounted silicon photomultipliers (SiPMs).

Crilin was optimised in the ambit of the Muon Collider experiment [1] as a candidate design for an electromagnetic barrel calorimeter, because of its fine granularity, excellent timing resolution, good pileup capability and high resistance to radiation. As verified from simulation, in the case of a Muon Collider, a barrel electromagnetic calorimeter with fine granularity ($10 \times 10$ mm$^2$ cells), 5-layer longitudinal segmentation and single-cell time resolution better than 80 ps for E$_{dep}$ >1 GeV would provide good rejection of the challenging beam-induced background. This background from muon decay products and their subsequent interactions is characterized by particles with low momentum (~1.8 MeV), displaced origin, and asynchronous time of arrival.

Because of its flexible architecture, the application of the Crilin design is possible in many different physics scenarios. The Crilin architecture has indeed also been adopted as a candidate for the development of the Small-Angle Calorimeter (SAC) for the HIKE experiment [2], for which a highly granular, longitudinally segmented, fast crystal calorimeter with SiPM readout was independently proposed. The HIKE SAC will need to withstand a very demanding high-rate environment with intense radiation fields, while guaranteeing superior pileup capabilities and very high detection efficiency for photons.

In autumn 2022, a Crilin prototype module (Proto-0), along with a prototype version of the front-end electronics system, was tested with single PbF$_2$ and PWO-UF crystals using a 120-GeV electron beam at the CERN SPS H2 beamline. These tests were focused on the measurement and optimisation of the time resolution, the study of the light transport and collection dynamics, and the validation of the readout chain.

## 1.1 Calorimeter prototype

The Crilin calorimeter prototype used for the beam test (Proto-0) was developed in the ambit of the Muon Collider experiment [3]. Proto-0 houses two 10×10×40 mm$^3$ crystals. The mechanical structure was realised via fused-deposition modelling in acrylonitrile styrene acrylate (ASA) with an overall size of 61×40×44 mm$^3$ (Figure 1).

The baseline choice of crystal for the Crilin calorimeter is lead fluoride, PbF$_2$. PbF$_2$ is a Cherenkov crystal [4] offering intrinsically fast emission, in line with the aforementioned stringent timing requirements. Alternative crystal choices are also under investigation, such as a recent formulation of lead tungstate with ultra-fast emission [5], now commercially available from Crytur [6] as PWO-UF. This material features high density, good light yield, high radiation resistance and fast response speed by combining the prompt Cherenkov emission with a fast scintillation component, yielding a dominant emission with a decay time $\tau < 0.7$ ns.

| Crystal | PbF$_2$ | PWO-UF |
|---|---|---|
| Density [g/cm$^3$] | 7.77 | 8.27 |
| Radiation length [cm] | 0.93 | 0.89 |
| Molière radius [cm] | 2.2 | 2.0 |
| Decay constant [ns] | - | 0.64 |
| Refractive index at 450 nm | 1.8 | 2.2 |
| Manufacturer | SICCAS | Crytur |

**Table 1.** Comparison of PbF$_2$ and PWO-UF crystals.

Both PbF$_2$ and PWO-UF crystals were employed during the beam test. Table 1 summarises the properties of these crystals. For the beam test, the crystals were wrapped in 100-$\mu$m-thick aluminized Mylar foil and tested one at a time in dedicated runs. Each crystal was readout by a 2×2 matrix of $3 \times 3$ mm$^2$ Hamamatsu

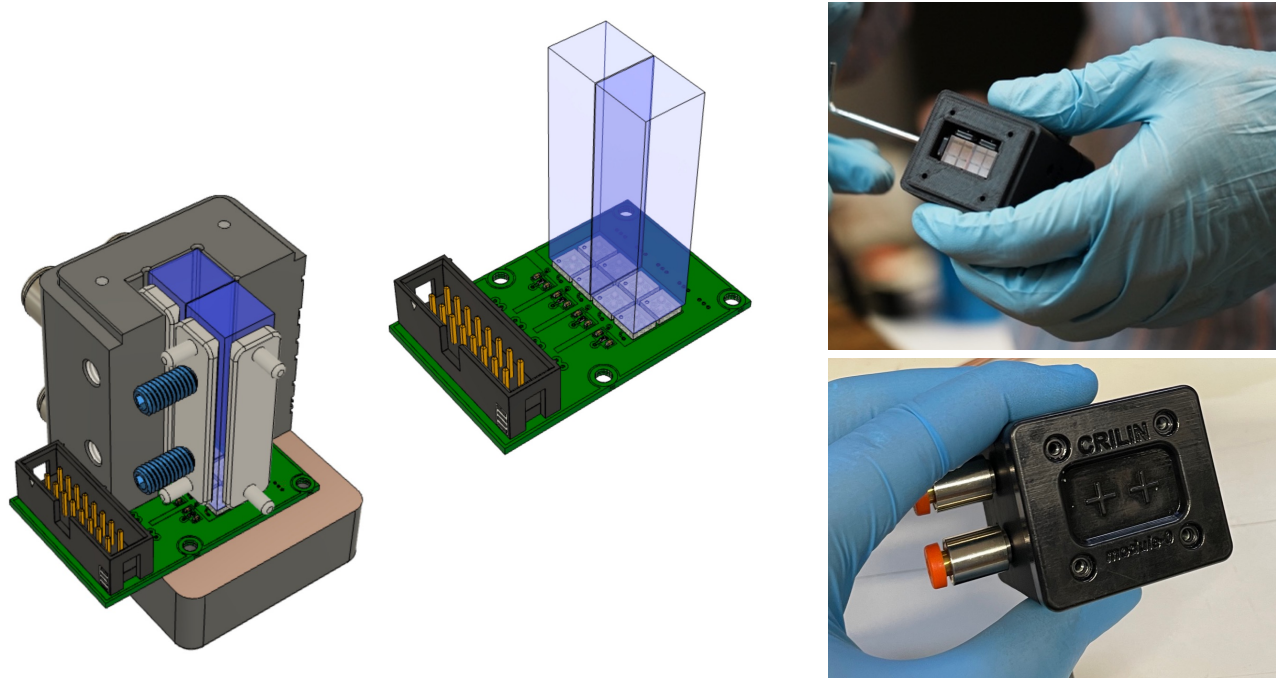

**Figure 1.** Left: Rendering of Proto-0 mechanics. Middle: Detail of Proto-0 SiPM board. Right: Pictures of Proto-0 during the assembly phase.

S14160-3010PS SMD silicon photomultipliers [7], with 10-$\mu$m pixel size, mounted on a dedicated PCB (SiPM board). The crystals were optically coupled to the SiPMs by direct contact without the use of optical grease. The left and right sides of the crystal were each read out by a pair of SiPMs connected in series, providing two independent readout channels for each crystal (see inset in Figure 3). The signals were transmitted from the SiPMs to the FEE board via micro-coaxial transmission lines. For the test, a two-channel prototype version of the Crilin front-end electronics (FEE) was used. On the FEE board, after proper termination, the pulses were processed first by a high-speed, non-inverting amplification stage with gain 4. The first stage output drove a pole-zero cancellation circuit, followed by a second, non-inverting stage (with gain 4) to drive the digitisation section. The FEE circuit has a dynamic range of 2 V and an overall gain of 8. External HV supplies were used for biasing.

## 2 BEAM TEST

### 2.1 Setup

All measurements were carried out in the H2 beamline at the CERN SPS with a 120-GeV electron beam and the setup illustrated in Figure 2. The trigger was obtained from the coincidence of the two scintillator counters, S1 and S2. The beam was tracked with a beam telescope consisting of two stations of two $9.5 \times 9.5$ cm$^2$ planes of silicon-microstrip tracking detectors, C1 and C2, spaced 15.4 m apart. Each single-sided tracking plane was 410 $\mu$m thick and had a spatial resolution of 47 $\mu$m, yielding an angular resolution for beam particles of about 3 $\mu$rad. Single particle event selection was performed by rejecting multi-cluster hits in the tracking detectors.

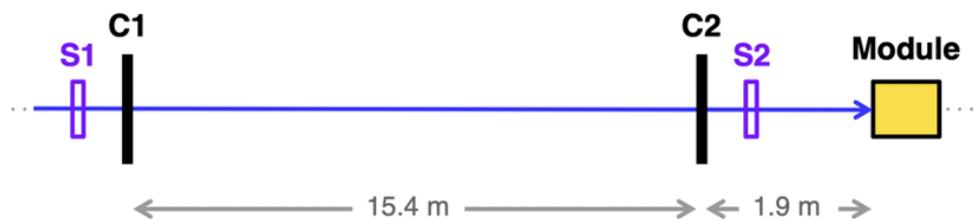

**Figure 2.** Top panel: Schematic representation of the beam test setup, scintillating counters (S1-S2) and tracking detectors (C1-C2), along with the positioning of the module under test.

The prototype was placed on a 4-axis motorised stage, with 2 axes of rotation and 2 axes of translation, for alignment (Figure 2, bottom). Two different data-taking configurations were employed:

- forward orientation, with the beam incident on the upstream face of the crystal and the SiPMs downstream facing upstream (front incidence);
- reversed orientation, with the beam incident on the back side of the SiPMs and the SiPMs at the upstream end of the crystal facing downstream (back incidence).

The tracking system made it possible to extrapolate the positions of beam particles at the crystal face, as shown in Figure 3, left. The *x-y* coordinate system has its origin at the center of the upstream crystal face, and the beam direction is anti-parallel to the *z* axis defined by right-hand coordinates. A detailed view of the readout geometry implemented on the SiPM board is shown in Figure 3, right.

### 2.2 Waveform reconstruction and analysis

SiPM signals from the two readout channels of Proto-0 were sampled at 5 GS/s using a CAEN V1742 switched-capacitor digitiser. For PWO-UF runs, as a consequence of the higher light yield, a 6 dB attenuator was placed before the digitiser inputs to halve the signal amplitude, due to the maximum 1 V dynamic range of the V1742 (in contrast to the 2 V output dynamic range of the FEE). The charge and amplitude values for PWO-UF reported in the text already account for the presence of the attenuator and are scaled to represent the true values output by the FEE system. The SiPM pairs in series were biased at 83.5 V, which corresponds to a 3.75 V overvoltage for each photosensor ($V_{br}$ = 38 V).

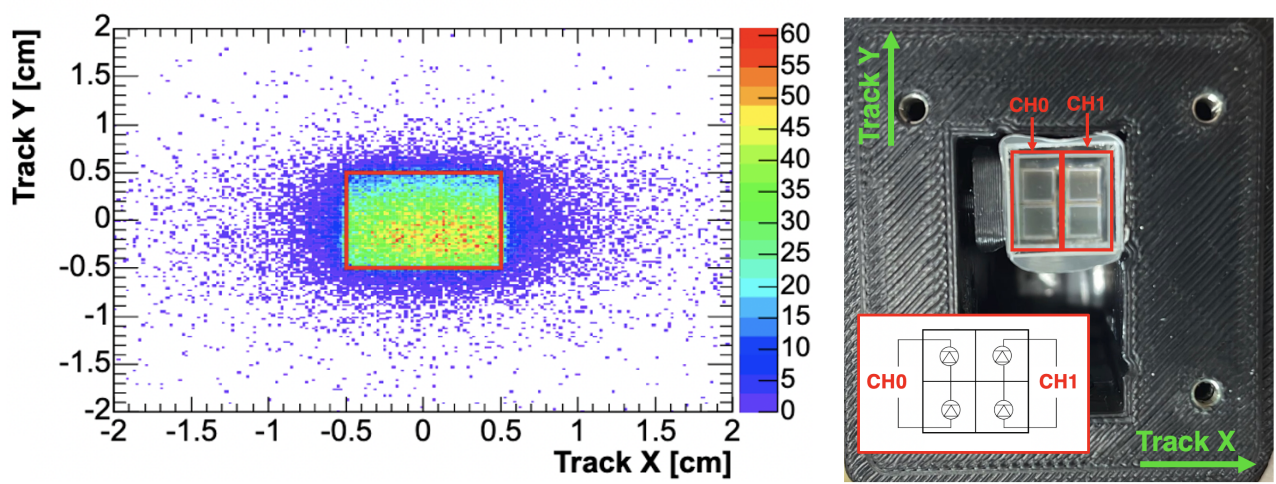

**Figure 3.** Left: Extrapolation of tracks to the upstream crystal face and localisation of the geometrical 1×1 cm$^2$ fiducial volume (red). Right: Photo of Proto-0 assembly. The PbF$_2$ crystal and SiPM matrix are visible (the front wrapping was removed). The SiPM series wiring scheme is shown in the inset and the tracking coordinate system in overlay.

Pulse charges were evaluated by integrating each waveform over the range [$T_{peak}$-20 ns, $T_{peak}$+140 ns], where $T_{peak}$ represents the waveform peak time, and dividing by the 50 $\Omega$ input impedance of the digitiser. Offline equalisation was carried out on the residual minor imbalances in charge between the two readout channels, less than 5% for PbF$_2$ and less than 2% for PWO-UF, due to small non-uniformities in SiPM gains and optical couplings. For the PbF$_2$ runs, events with signal of at least 50 pC on both readout channels (80 pC for PWO-UF) were selected, and fiducial cuts were made on the extrapolated position of the beam particle at the crystal face.

The pulse timing was evaluated using a waveform template fit procedure. SiPM pulse templates are sets of nodes with polynomial interpolation and fixed proportions, which can be fit to each waveform using a three-parameter optimisation. For each channel, waveform templates were generated by aligning and averaging signals from a large dataset of hits: for each sampled waveform, a pseudo-timing was extracted by applying a polynomial spline interpolation to the rising edge and peak, using a constant fraction technique (CF) applied to the spline function (Figure 4, left). The CF value employed for reconstruction was 12% of the peak amplitude, optimised by minimising the timing resolution, as shown in Figure 5 (top left). Finally, using the pseudo-timing information, all processed waveforms were aligned and, after proper normalisation, averaged into wave templates. A comparison between the waveform templates for PbF$_2$ and PWO-UF crystals is shown in the right panel of Figure 4: a sharper rising edge and narrower pulse shape is observed for PbF$_2$ due to different light generation and transport dynamics, as discussed later.

To reconstruct the pulse timing, templates were fitted to the rising edge using a three-parameter minimisation (scale, baseline and time offset) over the range [$T_{peak} - 20$ ns; $T_{peak} - 2$ ns]. An example of the application of the template fit is shown in Figure 4, left. The fit range bounds were optimised by minimising the timing resolution, as before. It should be noted that this reconstruction procedure does not introduce any significant time-amplitude slewing, so that no correction was necessary in data. The fitting procedure resulted in the normalised $\chi^2$ distribution for the fitted waveforms shown in Figure 5, bottom right. Pulses used for the analysis were required to have $\chi^2$ <30. It was verified that the cut on $\chi^2$ and choice of fit range did not introduce any significant bias with respect to particle hit position, as seen in Figure 5, bottom

*D. Paesani et al.*

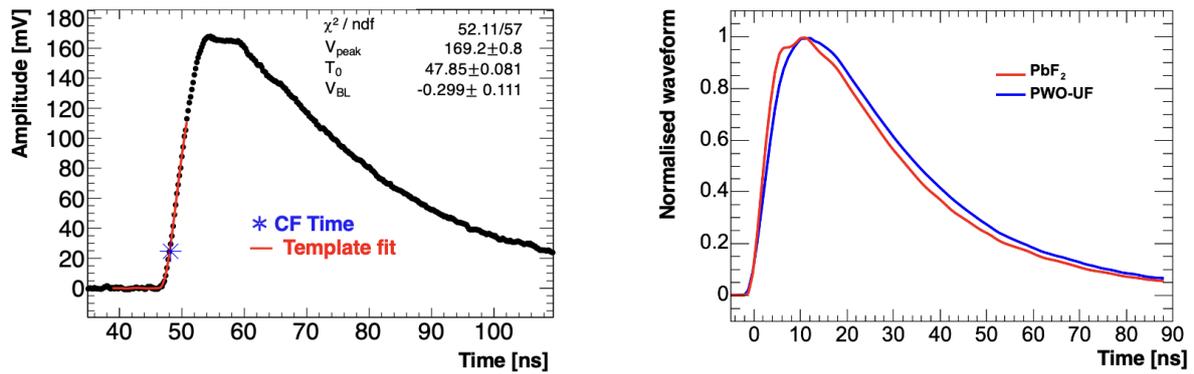

**Figure 4.** Left panel: Example of signal fitted using a template generated from corresponding dataset; the marker shows the constant fraction time. Right panel: Comparison between the pulse shapes for two different types of crystal. $PbF_2$ shows a sharper rising edge than PWO-UF.

left; such a bias might be expected due to the variation of the waveform shape as a function of particle hit position, which will be discussed in Section 4. Similarly, it was verified that the aforementioned selection cuts did not result in any bias with respect to pulse amplitude and timing. To ensure that the reconstructed timing information was free of any significant bias with respect to the digitiser sampling frequency, the plot in Figure 5, top right, was produced, which shows that the distribution of the reconstructed time modulo the digitiser sampling period is flat.

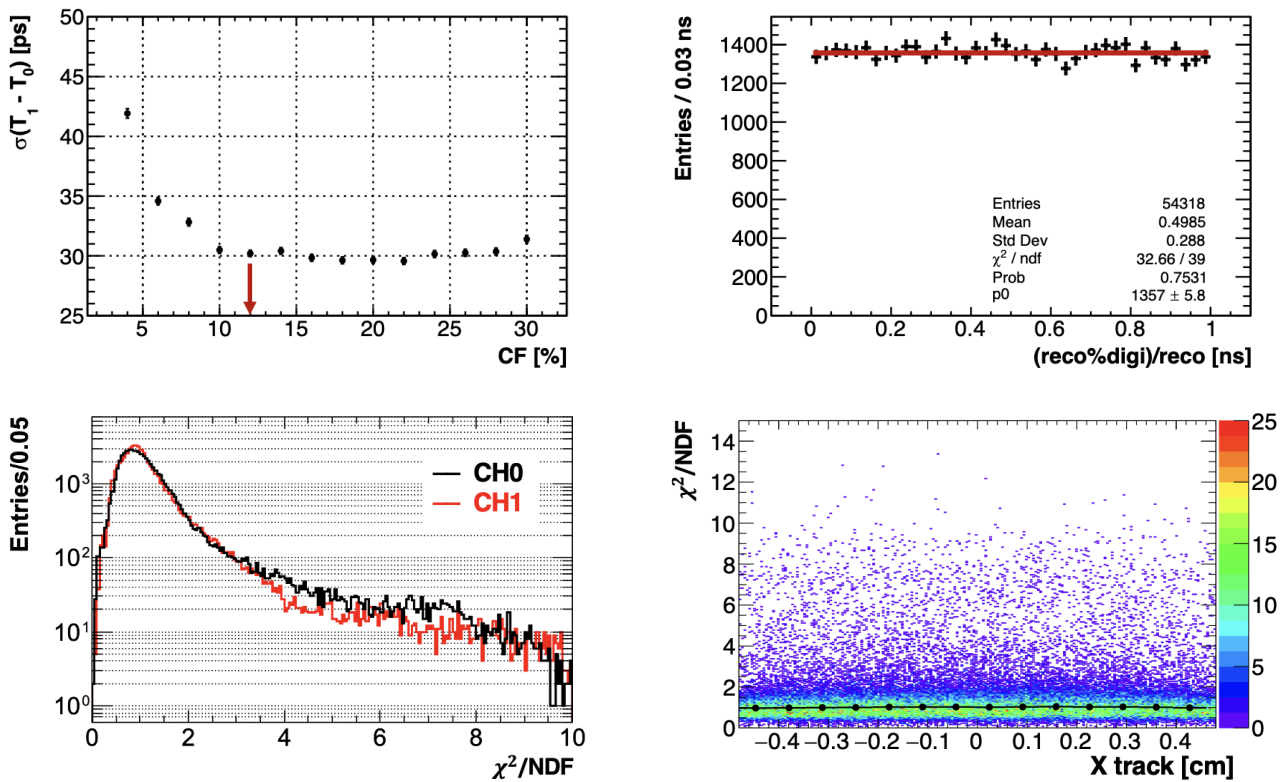

**Figure 5.** Timing reconstruction diagnostics. Top left: Example of constant fraction optimisation by minimisation of the timing resolution. Top right: Plot of reconstructed time modulo the digitiser sampling period of 200 ps, showing no significant bias from the timing algorithm. Bottom left: Example of $\chi^2$ distributions resulting from template fits applied to $PbF_2$ waveforms for CH0 (black) and CH1 (red). Bottom right: $\chi^2$ distribution as a function of the particle hit position on the crystal.

## 3 TEST BEAM SIMULATION

Detailed Geant4 [8] simulations of the beam, crystal, wrapping, and SiPM readout were developed for both types of crystals. A sensitive detector attached to the crystal volume was used to score energy deposits, while different beam sources were used to reproduce the test beam scenarios, as discussed below. Figure 6 shows the reference geometry, containing a single crystal and its wrapping, along with the four SiPM packages and active silicon regions.

### 3.1 Optical transport

For $PbF_2$ crystals, which represent the baseline choice for the Crilin design, a detailed simulation was also implemented to study the optical transport of Cherenkov photons. The relevant optical properties and surfaces were simulated. In particular, a dielectric-dielectric optical boundary between the $PbF_2$ crystal and Mylar wrapping was implemented, based on the LUT model [9]. The interface between the crystal and the four SiPM packages, made of silicone resin, was simulated using a polished dielectric-dielectric boundary (UNIFIED model). As shown in Figure 6, four 3×3 $mm^2$ active regions made of silicon were used to reproduce the active areas of the SiPMs. Sensitive detectors attached to the four silicon regions were used to score the energy, position, and timing of optical photon hits.

### 3.2 Digitisation

Optical photons arriving on the sensitive detector volumes representing the SiPMs for each readout channel were counted and used to simulate the corresponding signal waveform (Figure 6). To evaluate the number of detected photoelectrons, optical photon hits were weighted offline based on the spectral response of the photodetector, which has a peak PDE (photon detection efficiency) of 18% at 450 nm [7]).

For each simulated event and each readout channel, a SiPM pulse template representing the contribution of each individual pixel (single photoelectron response) was convoluted with the arrival times of optical photons over an interval of $[-1, +100]$ ns with respect to the particle hit time on the crystal surface. The resulting pseudo-waveforms were fitted using the template method discussed above (Section 2.2) to extract timing information.

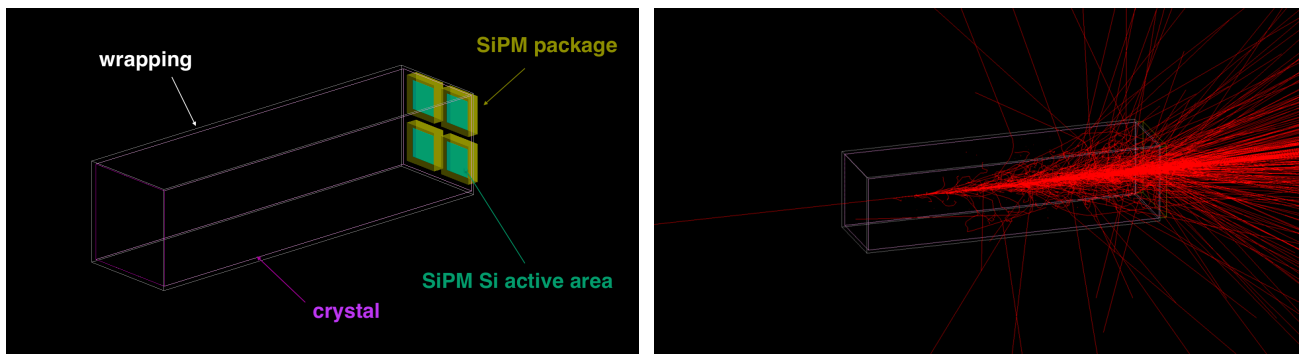

**Figure 6.** Left: Geometry of the Geant4 simulation. Right: Example of shower development for a 120-GeV electron incident on the center of the front face of the crystal (optical photon tracks are not shown).

# 4 RESULTS

## 4.1 Energy scale and light yield

To evaluate the prototype response in terms of output charge per unit of deposited energy, mean-charge distributions for events with the track incident within a square $5 \times 5$ mm$^2$ fiducial region centred on the front face of the crystal were compared to the analogous distributions from Geant4 simulations carried out using a planar, 120-GeV electron source of the same dimensions. For both types of crystals, a total of $10^5$ events were generated by resampling $x$-$y$ beam positions from the ones actually tracked during the test beam. The MC energy histogram showed a most probable energy deposit of about 4.9 GeV for PbF$_2$ and 5.8 GeV for PWO-UF. For all runs, the histogram of the deposited energy distribution from the MC was fitted to that for data using normalisation and scale parameters. From the fit procedure, the scale factors 29, 36, 67 and 77 pC/GeV were obtained for the cases of PbF$_2$ back, PbF$_2$ front, PWO-UF back, and PWO-UF front, respectively. After fitting, the data-MC consistency in shape was checked using a Kolmorogov-Smirnov test, resulting in a $p$-value $> 0.5$ for all PbF$_2$ runs and $> 0.3$ for all PWO-UF runs. An example of the data-MC overlay for PbF$_2$ is shown in Figure 7, where the range 50-300 pC was used for the MC shape fit. A comparison between the energy scale factors for PbF$_2$ and PWO-UF is shown in Table 7, along with the most probable values (MPV) and sigma values for the respective deposited energy distributions, as obtained from fits with a Landau distribution convoluted with a Gaussian resolution function.

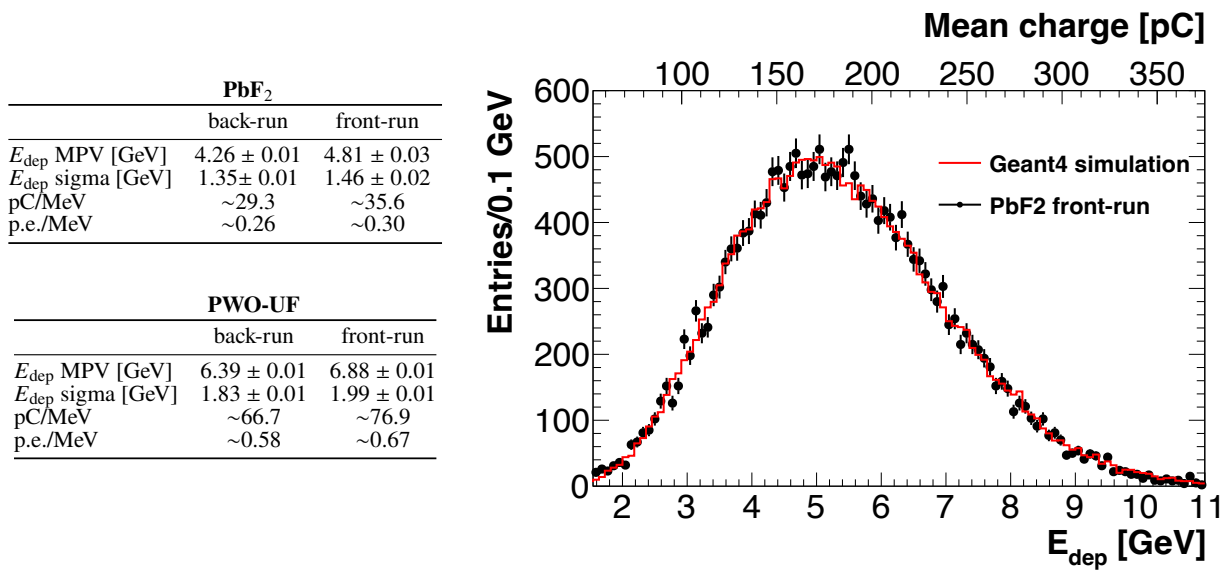

| PbF$_2$ | back-run | front-run |
|---|---|---|
| $E_{\text{dep}}$ MPV [GeV] | 4.26 ± 0.01 | 4.81 ± 0.03 |
| $E_{\text{dep}}$ sigma [GeV] | 1.35 ± 0.01 | 1.46 ± 0.02 |
| pC/MeV | ~29.3 | ~35.6 |
| p.e./MeV | ~0.26 | ~0.30 |

| PWO-UF | back-run | front-run |
|---|---|---|
| $E_{\text{dep}}$ MPV [GeV] | 6.39 ± 0.01 | 6.88 ± 0.01 |
| $E_{\text{dep}}$ sigma [GeV] | 1.83 ± 0.01 | 1.99 ± 0.01 |
| pC/MeV | ~66.7 | ~76.9 |
| p.e./MeV | ~0.58 | ~0.67 |

**Figure 7.** Determination of the energy scale. Right: example of data-MC overlay for energy deposit fit for PbF$_2$ front. Left: Summary of energy scale and relative parameters for the two crystals in both run configurations. The values of MPV($E_{\text{dep}}$) and $\sigma(E_{\text{dep}})$ were obtained via fits to a Gaussian-convoluted Landau distribution.

Once the scale factors have been determined, an estimate of the light yield can be derived from the knowledge of the SiPM gain. In particular, the SiPMs used have a nominal gain of $1.8 \times 10^5$ at $V_{\text{op}}$, as previously characterised [3]. Accounting for the charge gain of the amplifier in the FEE, the light yield values 0.26, 0.30, 0.58, 0.67 p.e./MeV were obtained for the cases of PbF$_2$ back, PbF$_2$ front, PWO-UF back, and PWO-UF front, respectively. For comparison, the simulation gives a light yield of 0.38 p.e./MeV for the case of PbF$_2$ in front configuration, after weighting according to the PDE of the SiPMs.

## 4.2 Light transport and position-dependent effects

For runs carried out in the front configuration, the waveform shape, along with the charge and timing distributions, presented some variation as a function of the particle hit position on the crystal. This behaviour is assumed to be associated with light transport effects inside the crystal that give rise to asymmetries in the light collected by the SiPMs for each of the two readout channels. These asymmetries are ultimately reflected in the apparent light yield and signal timing for each channel, as discussed in the following.

### 4.2.1 Modification of the waveform shape

A modification of the waveform shape as a function of the position of particle incidence is clearly visible for PbF$_2$, as already observed elsewhere [10]. The effect is also observed for PWO-UF, although it is less significant, possibly due to the presence of the isotropic scintillation component and the slower rise time (Figure 4, right). Figure 8 illustrates the pulse shape modification for PbF$_2$ as a function of the beam $x$ position. Normalised and aligned pulse profiles are shown for a single readout channel (CH0) when various fiducial cuts on the $x$ position of the incident particle are applied. Sharper rising edges are indeed associated with the more direct light component, which generally reaches the photosensor after few or no reflections and is characterised by earlier and sharper arrival times due to the directional nature of the Cherenkov light, as opposed to the indirect light component, which is delayed and spread in time by the multiple reflection modes and associated transit times inside the crystal. This latter effect is also associated with the slight but progressive broadening of the waveform as the location of beam incidence is shifted away from the active region of the photosensors.

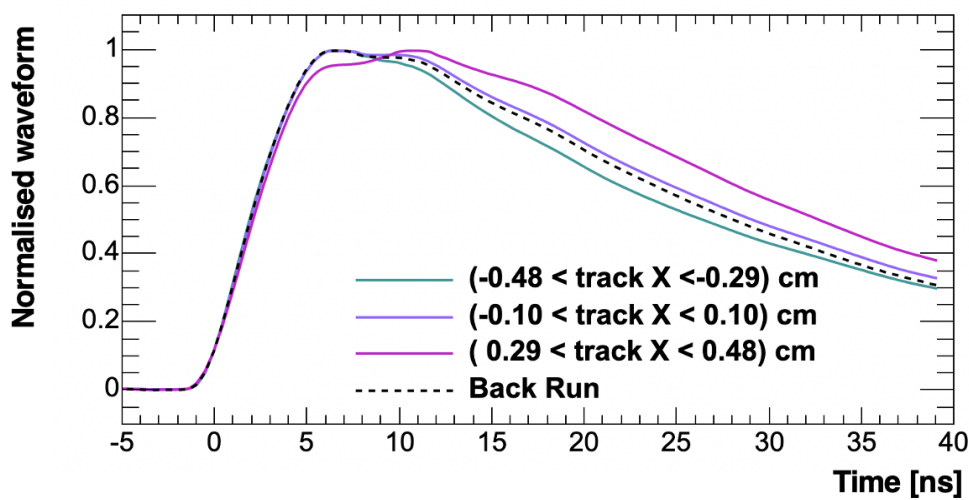

**Figure 8.** Example of pulse shape modification as a function of impact position selected with different fiducial cuts: green, for particle incident directly on SiPM pair giving signal; magenta, for particle incident on opposite SiPM pair; purple, particle incident between SiPM pairs. The dashed line shows the signal shape for back runs.

### 4.2.2 Effects on charge and timing

Figure 9 shows how the charge and timing distributions are also affected by the $x$ position of beam incidence, for the case of PbF$_2$. Figure 9, top, shows a plot of the asymmetry variable $A = (Q_1 - Q_0)/(Q_1 + Q_0) = (E_{\text{dep}\,0} - E_{\text{dep}\,1})/(E_{\text{dep}\,1} + E_{\text{dep}\,0})$ as a function of beam position in $x$, where $Q_0$ and $Q_1$ ($E_{\text{dep}\,0}$ and $E_{\text{dep}\,1}$) refer to the pulse charge (deposited energy) of the respective readout channels CH0 and CH1. The imbalance in charge between the two channels reaches its maximum ($\sim \pm 10\%$) when the beam is approximately centered on either of the two vertical SiPM arrays corresponding to the readout channels CH0 and CH1. Light propagated indirectly is more strongly attenuated due to the longer total path length traversed and the multiple reflections. The timing differences between the two channels as a function of the beam $x$ coordinate are also shown in Figure 9, bottom. As an intuitive consequence of the earlier arrival times for photons arriving directly, as discussed above, the charge and time differences between signals on the two channels are anti-correlated. As seen in the left panels of Figure 9, these asymmetries are not observed for the case of backwards incidence. This is because all of the Cherenkov light is emitted in the forward direction and must necessarily be reflected from the opposite end of the crystal before reaching the photosensors; the randomization of the trajectories washes out the correlation with the $x$ coordinate at which the light was originally produced. The same effects, resulting in similarly shaped distributions, were also observed in the case of PWO-UF, but with a larger charge separation ($\sim \pm 15\%$ maximum) and smaller timing separation ($\sim \pm 0.6$ ns maximum), possibly due to different light propagation dynamics arising from differences in optical parameters, wrapping and surfaces.

These inhomogeneities in light collection were also studied with the Geant4 optical simulation discussed in Section 3 to obtain a qualitative understanding of the light transport dynamics. Figure 10, top, shows the simulated spatial distribution of optical photons at incidence on the photosensor matrix in response to a 120-GeV electron beam with a 3-mm offset in the $x$ direction. As shown in Figure 10, bottom, the charge and timing asymmetries are correctly reproduced by the MC simulation when the beam source is scanned along the $x$ axis. Timing and charges were reconstructed from the simulated waveforms as described in Section 2.2. The charge profile is seen to be correctly reproduced, and the maximum asymmetry is compatible with the value observed in data to within 20%. For the timing profile, the shape is correctly reproduced, but the extent of the variation of the CH1-CH0 difference is significantly less for the simulation ($\pm 0.3$ ns) than for data ($\pm 0.5$ ns). This is probably due to the imperfect modelling of the optical surfaces in the simulation. Furthermore, the variation of the shape of the signal waveform as a function of the beam position in $x$ is not fully reproduced by the digitisation process of the simulation, due to differences in response of the readout chain, whose effects are not fully simulated. Despite these limitations, the fact that the simulation correctly reproduces the form of the charge-asymmetry and time-difference profiles, including their anticorrelation, demonstrates that the observed variations may be satisfactorily attributed to the light-transport effects described.

### 4.2.3 Comments and prospective improvements

It should be noted that similar but much less significant positional effects relative to beam shifts in the $y$ coordinate were observed, due to the geometry and series connection of the readout SiPMs. The discussion of this effect is beyond the scope of the current analysis, though alternative readout schemes (for example, parallel SiPM wiring) and their effects on timing performance are currently under investigation. Other measures under investigation to mitigate the position-dependent effects include the use of alternative, strongly diffusive surface treatments (for example, ground crystal surfaces).

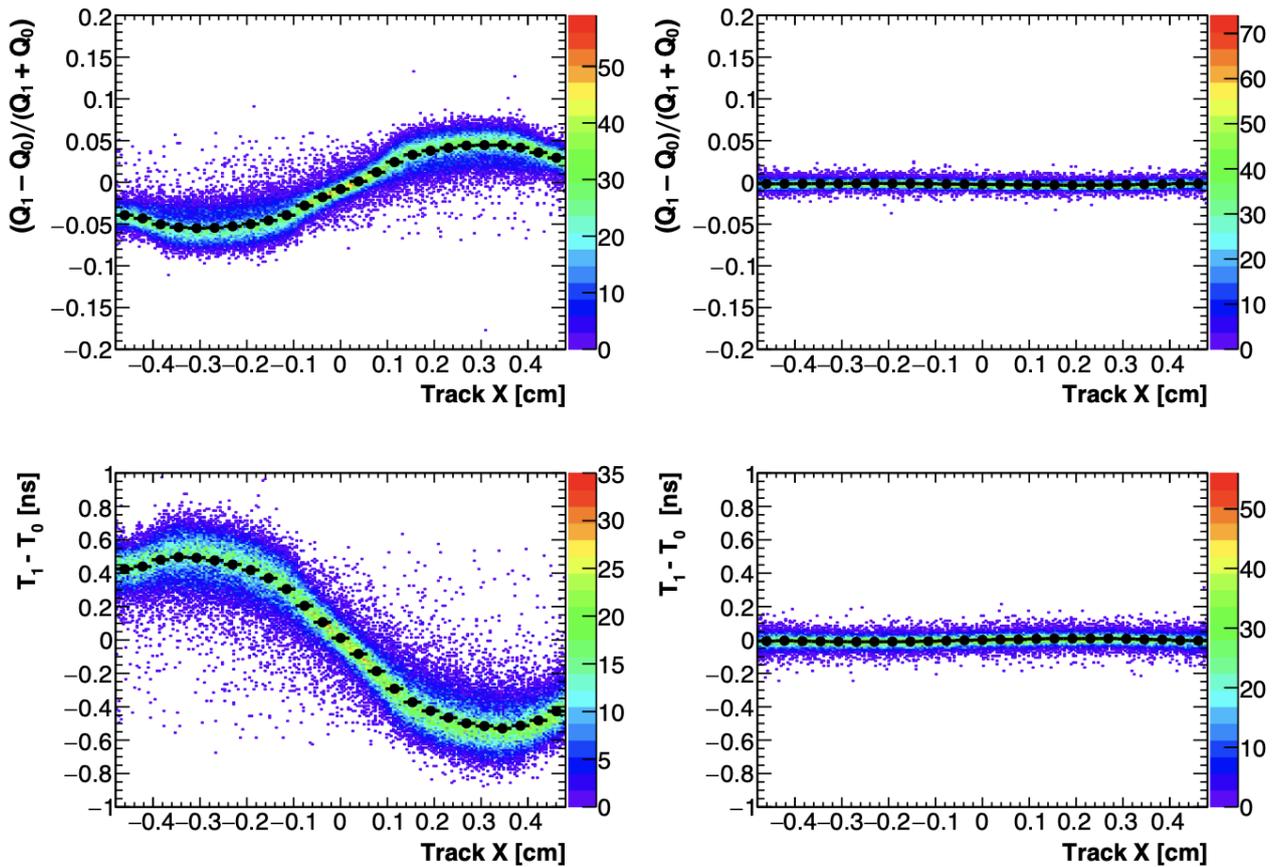

**Figure 9.** Top panels: Asymmetry variable $A = (Q_1 - Q_0)/(Q_1 + Q_0)$ as a function of the *x* position of beam incidence for $PbF_2$, front run (left) and back run (right). Bottom panels: timing differences between the two channels as a function of the *x* position of beam incidence for $PbF_2$, front run (left) and back run (right).

The position-dependent effects observed are particularly noticeable due to the small total longitudinal dimension ($\sim 4X_0$) of the single crystals under test. The early stages of shower development are characterised by a reduced number of secondary tracks with significant boost, so that the Cherenkov light emission remains strongly directional.

It should finally be noted that, in practice, the use of mean-charge and mean-time variables (with respect to the two readout channels) averages out all positional effects due to the light transport, as demonstrated, for example, by the mean-charge distributions in Figure 7 and the mean-charge and mean-time distributions in Figure 10, bottom. In the latter case, results from the simulation demonstrate that the reconstructed values of mean charge and mean time are completely independent of the position of beam incidence.

### 4.3 Timing performance

For all experimental configurations, the distribution of the time difference between the two readout channels $\Delta T = T_1 - T_0$ was used to study the time resolution of the system as a function of deposited energy. A $0.8 \times 0.8$ cm² fiducial cut centered on the crystal face was applied for all runs. The distribution of $\Delta T$ as a function of deposited energy $E_{\text{dep}}$ for $PbF_2$ runs is shown in Figure 11. The value of $E_{\text{dep}}$ is obtained from the mean of the charge values from both SiPMs, using the scaling factors discussed in Section 4.1. As shown in Figure 11, top left, for front-configuration runs, this distribution is split into two populations due to the position-dependent light transport effects described in Section 4.2. In order to evaluate the timing resolution in the front configuration, a correction was developed based on the dependence of $\Delta T$ on the charge asymmetry variable $A = (Q_1 - Q_0)/(Q_1 + Q_0)$, without relying directly on any information derived from the tracking system, as shown in Figure 11, top-right. Approximating the $\Delta T$-$A$ relationship with a straight line, a linear fit yields a slope of about $-8$ ns for PbF$_2$ (about $-4$ ns for PWO-UF).

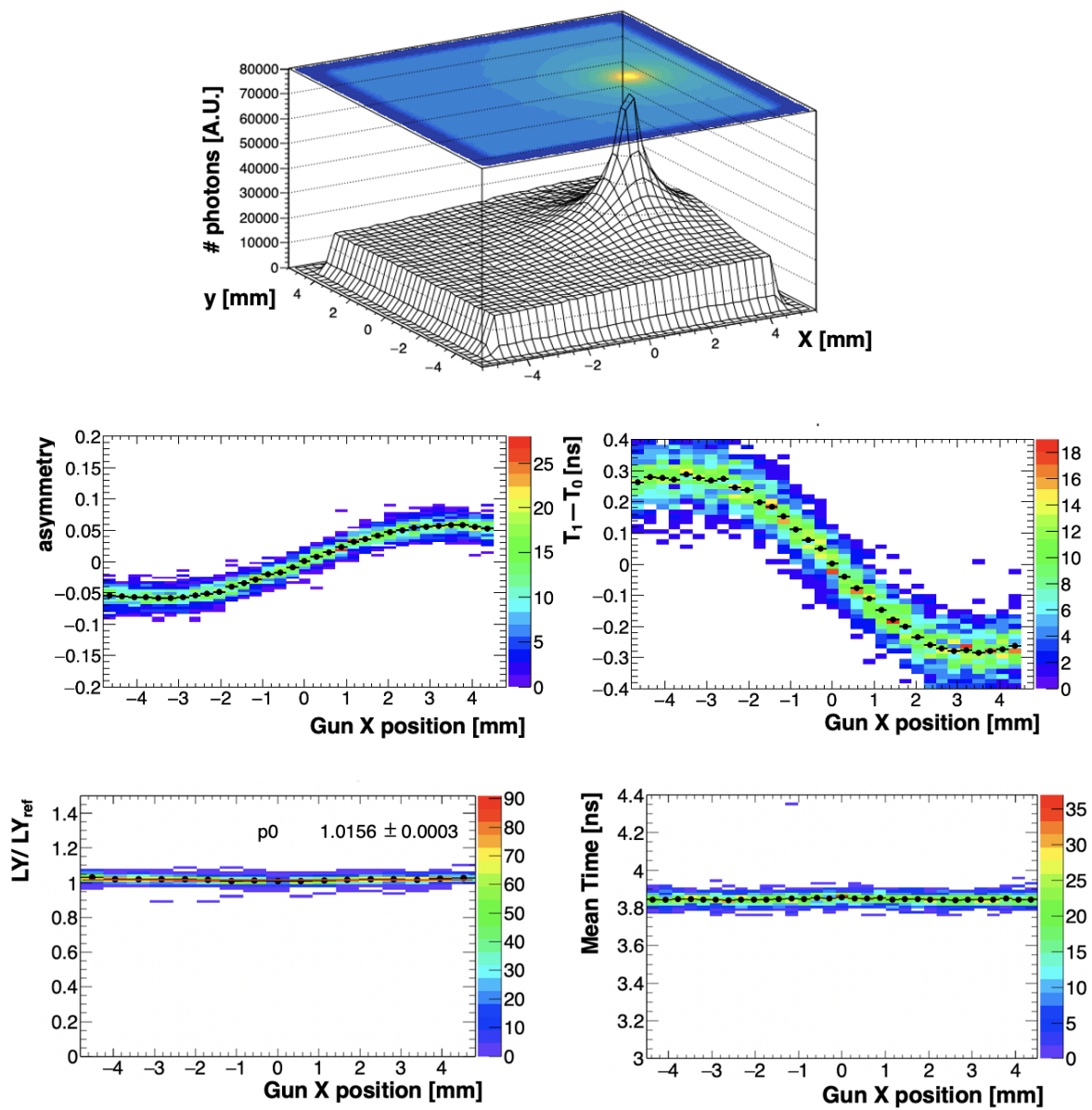

**Figure 10.** Results of the simulation. Top: spatial distribution of optical photons at the photosensor matrix from the interactions of 120-GeV electrons incident on the crystal with a 3-mm offset in the *x* direction. Middle: CH1-CH0 charge asymmetry and time differences as a function of beam *x* position when the 120-GeV electron beam is scanned along the *x* axis. Bottom-right: behaviour of the mean time for the two readout channels as a function of the beam position. Bottom-left: behaviour of the mean charge response as a function of the beam position, using the normalized light yield LY/LY$_{ref}$, where LY is the mean value of p.e./MeV for the two readout channels at a given beam position, and LY$_{ref}$ is the mean LY obtained with the centred beam. A constant function fit (p0) is overlaid.

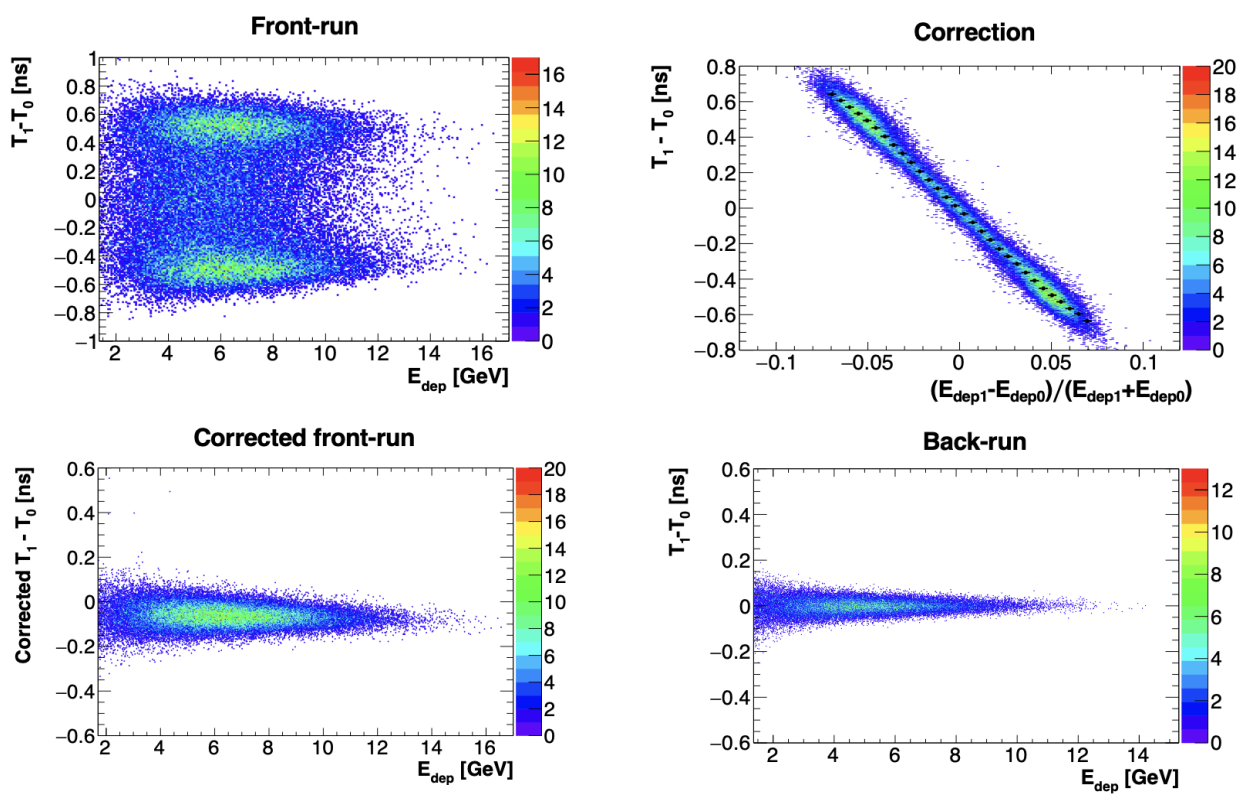

**Figure 11.** Overview of the procedure for the correction of position-dependent effects for evaluation of the timing performance. Top left: Time difference between the two readout channels as a function of $E_{\text{dep}}$, for front configuration runs. The splitting of the distribution from position-dependent effects is evident. Top right: Timing correction using charge asymmetry. Bottom: Distributions for runs in front (left) and back (right) configurations.

In order to correct the position dependence of the $\Delta T$ distribution, a spline function was fitted to the profile of the $\Delta T$-$A$ distribution and used to obtain an event-by-event correction for the timing offset due to positional effects. The corrected $\Delta T$ vs $E_{\text{dep}}$ histograms were then filled, as shown for PbF$_2$ in Figure 11, bottom left. The $\sigma_{T_1-T_0}$ time resolution could then be evaluated from the sigma of a Gaussian distribution fit to the corrected $\Delta T$ distributions for slices of $E_{\text{dep}}$. No correction was applied for runs carried out in back configuration. The results are summarised in Figure 12 for all runs, where the time resolution as a function of deposited energy was fitted using the function $\sigma_{\text{MT}} = \sigma_{T_1-T_0}/2 = a/E_{\text{dep}} \oplus b$, where the subscript MT refers to the resolution obtained for the mean time for the two readout channels of the single calorimeter cell. Even after the correction, the residual position-dependent light-transport effects spoil the timing performance for the front configuration, despite the generally higher light yields. Due to the purely Cherenkov nature of the light emission from PbF$_2$ for both configurations, the time resolution for PbF$_2$ is better than that for PWO-UF, despite the fact that the light yield for PbF$_2$ is only about half of that for PWO-UF.

For runs in the front configuration, it should be noted that, to correctly account for the charge imbalance between the two readout channels (in the worst-case, $\pm 8\%$ for PbF$_2$), the time resolution should ideally be modelled as $2\sigma_{\text{MT}} = \sigma_T(E_0) \oplus \sigma_T(E_1)$ instead of $2\sigma_{\text{MT}} = \sigma_{T_1-T_0}(E)$, where the subscript MT refers to the mean-time resolution, $E \equiv (E_0 + E_1)/2$, and $\sigma_T = \sigma_{T_0} = \sigma_{T_1}$ represents the resolution as a function of energy of a single readout channel (assumed to be identical). By generating trial time distributions in a toy-MC simulation assuming $\sigma_T(E) = \sigma_{T_1-T_0}(E)/\sqrt{2}$, it was verified that the worst-case discrepancy in the fit parameters obtained with the $\sigma_T(E)$ fit model is O(1%) for either PbF$_2$ or PWO-UF in the energy range of interest.

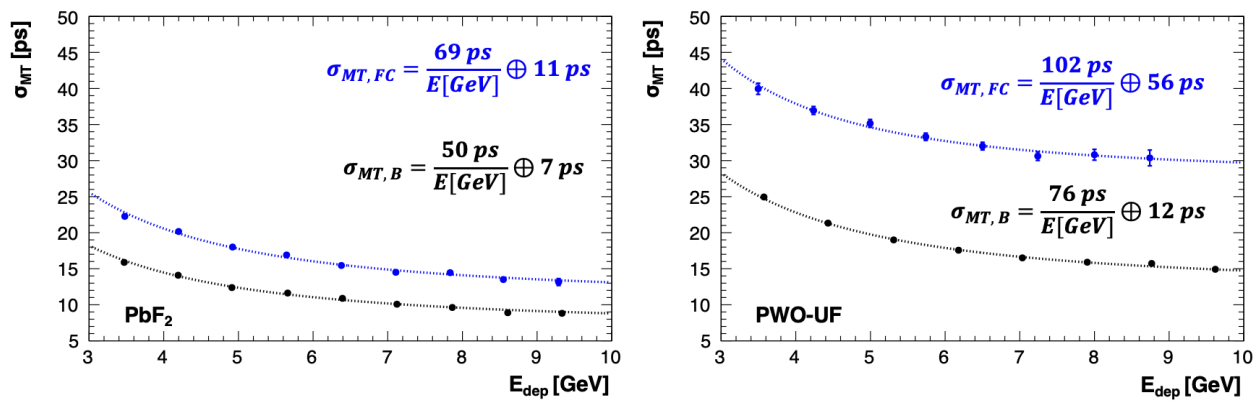

**Figure 12.** Mean-time resolution of a single calorimeter cell for PbF$_2$ (left) and PWO-UF (right) as a function of $E_{\text{dep}}$ over the range 3-10 GeV. Front-configuration corrected runs ($\sigma_{MT,FC}$) are shown in blue, while back-configuration runs ($\sigma_{MT,B}$) are shown in black.

## 5 CONCLUSIONS

Experimental progress in high-energy physics continues to demand modern and innovative solutions for high-performance, ultra-fast electromagnetic calorimetry.

Crilin is a promising design concept for a semi-homogeneous crystal calorimeter with longitudinal segmentation and SiPM readout, as demonstrated by the studies of small PbF$_2$ and PWO-UF crystals for use in the Crilin design described in this work. For a single $10 \times 10 \times 40$ mm$^3$ calorimeter cell of PbF$_2$, a worst-case time resolution (mean time of two SiPM readout channels) better than 25 ps (20 ps) is obtained for $E_{\text{dep}} > 3$ GeV, when the beam is incident on the front (back) face of the crystal. For a single cell of PWO-UF, a time resolution of better than 45 ps (30 ps) is obtained for this range of deposited energy. This timing performance fully satisfies the design requirements for the Muon Collider [11] and HIKE [2] experiments, while further optimizations of the readout scheme and crystal surface preparation may yet bring further improvements.

A more advanced Crilin prototype (Proto-1), consisting of two layers of 3×3 crystal matrices (for a total of 36 readout channels), was developed in 2022; a beam test campaign for its characterization is planned in 2023.

## ACKNOWLEDGEMENTS

This project has received funding from the European Union's Horizon 2020 Research and Innovation program under the AIDAinnova project, grant no. 101004761, as well as from INFN projects RD_MUCOL and NA62. The authors thank E. Auffray, L. Bandiera, D. De Salvador, D. Lucchesi, and N. Pastrone for their support, and Yu. Guz for the loan of the MCP-PMTs used for some of the timing studies.

## CONFLICT OF INTEREST STATEMENT

The authors declare that the research was conducted in the absence of any commercial or financial relationships that could be construed as a potential conflict of interest.